# East Asian Observations of Low Latitude Aurora during the Carrington Magnetic Storm

Hisashi Hayakawa (1), Kiyomi Iwahashi (2), Harufumi Tamazawa (3), Hiroaki Isobe (4, 5), Ryuho Kataoka (6, 7), Yusuke Ebihara (5, 8), Hiroko Miyahara (9), Akito Davis Kawamura (3), Kazunari Shibata (3,5),

(1) Graduate School of Letters, Kyoto University, Kyoto, Japan
(2) National Institute of Japanese Literature, Japan
(3) Kwasan Observatory, Kyoto University, Kyoto, Japan
(4) Graduate School of Advanced Integrated Studies in Human Survivability, Kyoto University, Kyoto, Japan
(5) Unit of Synergetic Studies for Space, Kyoto University, Kyoto, Japan
(6) National Institute of Polar Research, Tachikawa, Japan
(7) Department of Polar Science, SOKENDAI, Tachikawa, Japan
(8) Research Institute for Sustainable Humanosphere, Kyoto University, Kyoto, Japan
(9) Musashino Art University, Tokyo, Japan

**Abstract**

The magnetic storm around 1859 September 2, caused by so-called Carrington flare, was the most intense in the history of modern scientific observations, and hence is considered to be the benchmark event for space weather. The magnetic storm caused worldwide observations of auroras even at very low latitudes such as Hawaii, Panama, or Santiago, and the available magnetic field measurement at Bombay, India, showed two peaks: the main was the Carrington event which occurred in day time in East Asia, and a second storm after the Carrington event which occurred at night in East Asia. In this paper, we present a result from surveys of aurora records in East Asia, which provides new information of the aurora activity of this important event. We found some new East Asian records of low latitude aurora observations caused by the storm which occurred after the Carrington event. The size of the aurora belt of the second peak of the Carrington magnetic storm was even wider than usual low-latitude aurora events.

## 1. Introduction

The geomagnetic storm associated with the so-called Carrington solar flare observed in 1859 by Carrington (1859) is believed to be the most intense geomagnetic storm in the history of modern scientific observation (Tsurutani et al. 2003; Lakhina et al. 2013; Lakhina et al. 2016). It took only 17 h for the coronal mass ejection (CME) to travel from the sun to the earth, from which the CME speed is estimated to be about 2300 km/s. The intensity determined by Tsurutani et al. (2003) was Dst = -1760 nT, based on calculations using the aurora observations reported in Kimball (1960). This well-known event is regarded as the benchmark for extreme space weather events, and has attracted the attention of many researchers (Tsurutani et al. 2003; Cliver et al. 2004; Clark 2007).

From a few days before the Carrington flare, the sunspot group became active and caused extensive auroras observed all over the world (Kimball 1960). Especially just after the observation of the Carrington flare, auroras were observed simultaneously in North America and South America; and even in the places at low latitudes such as Guadeloupe, The Bahamas, and Honolulu (Loomis 1859; 1860a; 1860b; 1860c; 1860d; 1861a; 1861b; 1865). This magnetic storm even damaged telegraphs and caused fires with spark discharges (Loomis 1859; 1860a; 1860b; 1860c; 1860d; 1861a; 1861b; 1865; Green & Boardsen 2006). We can state that this is the first known case of an extreme space weather hazard. According to an estimate, if a magnetic storm as large as the Carrington event hit the earth today, global civilization would suffer from as much as USD 2 trillion dollars in damage (National Research Council 2008).

We do not know exactly how much energy was involved in the Carrington event (either for the solar corona, the coronal mass ejection (CME) or the associated energetic particles) (Lakhina et al. 2012; Tsurutani and Lakhina 2014). The size of a sunspot is an indirect indicator of the maximum energy available for the flare. The sketch of the sunspot drawn by Carrington (Figure 1, Figure 2) indicates that, by rough estimation, the total sunspot area is about 3000 mhs (millionths of a hemisphere). . Then, using equation (1) in Shibata et al. (2013) the upper limit of flare energy can be estimated to be about $4 \times 10^{33}$ ergs.

Another indicator of the strength of the event is the geomagnetic latitude of the attendant auroras. Most of the reports cited in previous studies of observations of the low-latitude auroras caused by the Carrington event focused on reports in North America, South America, or Europe. Kimball (1960) catalogued all the observational sites known in his time with geographical coordinate and geomagnetic latitude. On the other hand, from the Eastern Hemisphere, we have few reports for the contemporary low-latitude auroras. Chapman (1957) reported that aurora were seen in India. For East Asia, in particular, there was only one observation from Japan and one from China (Kimball 1960; Nakazawa et al. 2005; Willis et al. 2007; see next section for more detail). Thus, East Asia aurora observations associated with the Carrington magnetic storm event have not been reported.

Green et al. (2006) related this lack of records in this area, along with the contemporary weather, based on records (RG24 and RG27) of the American Navy in the National Archives and Record Administration (NARA).

In this study, we made a comprehensive survey of the records related to the Carrington event in East Asia: Japan, Korea, and China. In order to examine the cause of the relatively small number of aurora observations in East Asia, we also compile the records of weather at various locations in Japan and Korea. We also provide the information of the source documents in order to evaluate the reliability of the records.

## 2. Methods
### 2.1 Review of the previous works on East Asian records

There are two records of East Asian observations of the aurora associated with the Carrington magnetic storm that have been reported previously in the scientific literature: one from Singu (Wakayama) in Japan (record J1 in the list below) and one from *Luánchéng*, China (record C1 in the list below). The latter is originally catalogued by the Beijing observatory (1988) and introduced by Willis et al. (2007) with an English translation of the original text. Regarding the former, there seems to be a complicated history of citation and resultant confusion of the location of the observation. Here we summarize the citation history.

It was Musha (1939) who introduced the aurora observation in *Shingu* to the Japanese scientific community. Musha (1939) did not provide the detail bibliographic information of the source document and only mentioned that the record was taken from *Koutei nendai ki*, from *Singu, Kii*, where *Singu* is the name of the city which is still being used, while *Kii* is the old name for the region that mostly overlaps with Wakayama prefecture in present Japan. However, the original text of the *Koutei Nendaiki,* which usually refers to kind of documents related with *Kumano Nendaiki*, has not been identified as far as we know. It seems Musha (1939) took the record from *Singu shishi* (1937), the official history compiled by *Singu* city, which has a chapter of *Koutei Nendaiki* that includes the auroral record. The original text provided below is based on the copy of *Shingu shishi.*

Musha's report was included in the Japanese catalogue issued by the Central Meteorological Observatory (1943), in which the observational location was written as *Kii*, although no original text was given. Then the CMS's catalogue was cited in Matsushita (1956) who introduced this record for the first time in English (or in any other non-Japanese languages). Unfortunately, Matsushita (1956) described the location of the observation as *Wakayama*. Confusion arose here, because *Singu* city is located in the *Kii* region (approximately present Wakayama prefecture), but there is also Wakayama City (N34°.13′, E135°10′) in Wakayama prefecture, which is located about 90 km northwest of Singu city (N33°44′, E135°59′). Even more confusing, the location of "Wakayama" is indicated at the location of present Wakayama city in Figure 1 of Matsushita (1956). This mistake seems to have been transmitted to Kimball (1960) in which Matsushita (1956) cites the location of the observation is described as Wakayama, N34E135.

Kanda (1963) and Osaki (1994) also mentioned the same record, although both of them include only the original text in Japanese. Nakazawa et al. (2004) cited Osaki (1994)[1] and wrote the location of the observation as "Kii (Wakayama)" and the latitude is given as N34°2′, i.e., that of Wakayama city (no longitude information is given). Green and Boardsen (2006) cited Kimball (1960) and Nakazawa et al. (2004) in their paper that compiled the reports of the Carrington aurora from all over the world. Note that, however, in Green and Boardsen (2006)'s figure (Figure 2) that shows the geographical locations of the aurora observations, there is only one point in the sea between Japan and the Korean Peninsula. Finally, Willis et al. (2007) also introduced this record by citing Osaki (1994) with mention to the name of original text (*Koutei Nendaiki*) and the location is correctly given as `Shingu, Wakayama prefecture (33.43N, 136.00E)`.

Recently, Watanabe (2007) privately published a new catalogue of historical astronomical records in Japan. Although his catalogue has significant amount of transcription errors, it includes previously unknown records (J2, J3-3, J4). These records are included in this paper, but instead of translating the catalogue of Watanabe (2007) into English, we have read and the primary sources and included their bibliographic information.

**2.2 Source documents**

First, we re-examined the documents that include previously reported records of the Carrington aurora event. Here we emphasize that we re-examined all the primary sources as far as we could track. The nature and the reliability of the source documents are discussed in section 3.1. In addition, we have surveyed the descriptions of likely auroras or any anomalous celestial phenomena observed during 1859/08/28 to 1859/09/04 in other sources: one from China, two from Korea and many Japan. All the surveyed/re-examined documents are listed below. During the survey/reexamination, we have also searched for the information of the weather in order to examine the sky condition.

---

[1] Note that Osaki is spelled as Ohsaki in the reference list of Nakazawa et al. (2004).

In the case of book publications, the format of the document list is: ABBREVIATION: Author Name, *Document Name,* Place of publication, Year. In the case of handwritten manuscripts, it is: ABBREVIATION: Author Name, *Document Name,* reference number, owner. Inside the parenthesis is corresponding Chinese characters. Comments follow in any.

**Chinese documents.**

LZ: Chén Yǒng (陳詠), Zhāng Dūndé (張惇德), *Luánchéngxiànzhì* (欒城縣志), Luánchéng, 1872−74: 史-XI-4-A-419, Institute for Research in Humanities, Kyoto University., This includes the aurora record C1 reported by Beijing observatory (1988) and Willis et al. (2007).

QQ: Anonymous, *Qīngdài Qǐjūzhù in Xiánfēng Dynasty* (清代起居注 咸豐朝): I-LVII, Taipei, 1983. This is the source document for the official history of *Qīngshǐgǎo* (清史稿), whose likely aurora records were surveyed by Kawamura et al. (in press). Eventually no likely aurora records around Carrington event was found from this document.

**Korean documents**

ISN: Classic Publishers at Seoul National University (ed.), *Ilseongnok* (日省錄), Gwangmyeong, 1991.

SCWIG: Committee of National Historiography, *Seungcheonwon Ilgi* (承天院日記), Seoul, 1961.

These two Korean documents are governmental diaries that include the records of celestial records that were also used in Lee et al. (2004). Eventually no aurora-like records around Carrington event were found from these Korean documents.

**Japanese documents**

YKM: Yamauch Kanagiya Matasaburo, *Yamaichi Kanagiya Matasaburo Nikki* (山一金木屋又三郎日記), YK215-19-15, Hirosaki City Library. This document was included in the catalogue of Watanabe (2007), although we have found new aurora records (J3-1, J3-2) that were not included in Watanabe (2007).

YUS: Yorioka Ubei, *Yorioka Ubei Shojihikae* (依岡宇兵衛諸事控), taken from the History of In'nan Town （印南町史）, 1987. This document and its aurora record (J2) were included in Watanabe (2007).

KTN: Anonymous, *Kotei Nendaiki*（校定年代記）, taken from the History of Shingu City (1937),

p1216. This is the document that includes the Japanese record (J1) previously cited by many authors as mentioned above.

KNT: Anonymous, *Kenbun Nennen Tebikae*（見聞年々手控）, taken from the Local Hisotry of Hiraka Town (1969), p315.（平鹿町郷土誌』, 1969. This document and its aurora record (J4) is included in Watanabe (2007).

The following Japanese documents did not include likely aurora records, although some of them provided information of local weather.

JW01: Fukuma Y (2010)[2]

JW02: Fukuma Y (2014)

JW03: Hirosaki Han Okuni Nikki (弘前藩御国日記), Hirosaki City Library, TK215-1

JW04: Nichiyo Zakki (日用雑記), 30J/2049, National Institute of Japanese Literature.

JW05: Nikki (日記), 25C/281, National Institute of Japanese Literature.

JW06: Nikkan (日鑑), 24M/2392, National Institute of Japanese Literature.

JW07: Tomegakinarabi Nikki (留書并日記), 32E/15, National Institute of Japanese Literature.

JW08: Nichisakki (日刧記), 44G/108-2, National Institute of Japanese Literature.

JW09: Nenchu Yorodzu Nikki (年中万日記), 27D/1085-7, National Institute of Japanese Literature

JW10: Sekiguchi Nikki (関口日記) in Yokohama City (1979: XIV, pp114-115)

JW11: Ishikawa Nikki (石川日記) in Hachioji City (1990: XII, p25)

JW12: Nikki (日記) Kawasaki City Archive, facsimile version, Nr. 9.

JW13: Moriya Toneri Nikki (守屋舎人日記) in Moriya (1987: VIII, pp196-198)

JW14: Higashikuze Michitomi Nikki (東久世通禧日記) in Higashikuze (1992: I, pp266-268)

JW15: Onchu Nichinichirokunokoto (穏中日々録酒事), 864, Hachioji Local Museum.

JW16: Tanaka G (1990) Okyo Zakki (應響雜記), Toyama.

## 3. Results
### 3.1 Aurora Observations during 1859/08/28−09/04

As also commented in the previous section, in total we found one record in China and six aurora records in Japan that look like aurora observations, but none in Korea. Interestingly, the official histories in China or Korea such as QQ (see, Kawamura et al. 2015), ISN (pp834−847), or SCWIG (pp397–408) did not mention any aurora-like event around that time, and it was the local histories and private diaries in China and Japan that had what we searched for. Below we provide the list of the likely aurora records. The format in the first like is ABBREVIATION of record, ABBREVIATION of the source, page or figure numbers when available. It is followed by the original text, English translation by the authors, and (likely) location of the observation. The information such as color, direction and time is summarized in Table 1.

---

[2] JW01 and JW02 are extracts from JW03.

**Chinese records**

**C1:** LZ III, 19b (Figure 3)

**Original Text:** 清官咸豐九年…秋八月癸卯夜，赤氣起於西北，亘於東北，平明如滅。

**Translation:** On the night of 1859/09/02, a red vapor appeared from the west-north and moved toward the northeast. It disappeared at dawn.

**Location:** *Luánchéng* (欒城縣), *Héběi*, China (N37°53′, E114°39′)

**Japanese records**

**J1:** KTN, pp1216

**Original Text:** 八月六日夜六ツ時より夜半に及び，北方火災の如く紅し，明和七年七月二十八日，赤氣北方に現はれしことあり，それより九十年目になる

**Translation:** At night, from *Mutsudoki* (17:00-19:00) until midnight, it was as red as a conflagration in the north. This is the first time in 90 years since red vapor had also appeared in the north on 1770/09/17.

**Location:** *Shingu* (新宮), *Wakayama*, Japan (N 33°50′, E 135°46′)

**J2:** YUS, pp796

**Original Text:** 八月六日七ツ頃より北方雲赤くク，四ツ過迄不思議の事，若山湊辺出火之よし，凡千軒ほと焼候由噂御座候，同所ニテハ無之兵庫西宮之辺之酒屋と申事，同所ニテハ丹波辺と申事，当村ニテも高野山とやら，五条橋本と申すものも有之，中々火事ニテハ無之，金雲之氣先ツ変也

**Translation:** On 1859/09/02, from *Nanatsudoki* (15:00-17:00), clouds in the north were red, and the mysterious thing continued until *Yotsu* (21:00-23:00). It was rumored that a conflagration had broken out at *Wakayama Minato* and around a thousand houses had burnt down. Other rumors indicated that it was not at *Wakayama Minato,* but rather in liquor stores at *Hyogo Nishinomiya*. Over there, it was rumored that there was a conflagration at *Tamba.* In my village, there were rumors of fires at *Koyasan*, while other rumors indicated that the fires were at *Gojo Hashimoto*. It does not seem to be a conflagration. After all, a cloud of golden vapor is strange.

**Location:** *In'nan* (印南), *Wakayama*, Japan (N 33°59′, E 135°13′)

**J3-1:** YKM，安政６年８月６日条 (Figure 4)

**Original Text:** （安政八年八月）六日明け少冷敷薄黒く曇り西北より東北迄明き薄赤し

**Translation:** On the 6$^{th}$, it was a little cold, gloomy, and cloudy at dawn. From the west-north to the east-north, it was bright and faintly reddish.

**J3-2:** YKM，安政６年８月６日条 (Figure 5)

**Original Text:** 悴御寺へ用の参り暮過帰宅す、弘前より途中ニ而見し由、北の方赤く火の出候様に見候由、出火ニ可有之哉、不相分候而寝込出火ニ有之義哉と困り候而も、屋根登り見し由

**Translation:** In the evening (a little past 18:00), my son returned home from the temple. He said that on his way home, the northern sky seemed to be burning. Did a fire break out? It would be dangerous if I fell asleep without realizing there was a fire. I climbed up on a roof to look at the sky.

**Location:** *Hirosaki* (弘前), *Aomori*, Japan (N 40°36′, E 140°28′)

**J3-3:** YKM，安政６年８月９日条 (Figure 6)

**Original Text:** 六日夜北之方に見え候赤み、青森にても向こう地出火と見候よし、又は大畑辺、平内在かと皆下へ参り見候よし、しかる処夜九ツ時頃見候ところ青森向こうの方明るく有りし候よし、茂森うし朔日の夜中青森へ参り、一昨日帰宅仕り候話なり、尤も向こう地出火なれば一両日の内にも便り候よし、またこの明かり全く火事に有らず、この日の暮れ誠に西北・西南南迄赤き事珍しき雲立ちなり、依って夕焼け残居候と申す人も有り候へども、夕焼けは夜迄見え候義これまた不思議なり、弘前藩や話なれば先年昔大地震の前に北方大いに赤き事これ有り候、火事に無く相見え候事有り候ノよし、又オロシヤ大火事に有るべきや、又は奥地ソウヤ・カラフトなどの出火かなどと話有り候よし、なんだか相分からざる事、拙は暮れに南の夕焼けばかり見候、当年珍しき赤み有り候

**Translation:** The redness seen at night on the 6$^{th}$ seemed to be regarded as a fire. Everyone came out to see if it broke out at *Ohata*, or if it broke out in the countryside of *Hirauchi*. However, when I saw it at 23:00-01:00, it got brighter near Aomori as well. This story was told by Mr. *Shigemori*, who departed to *Aomori* on 08/28 and came back here the day before yesterday. After all, if a fire broke out over there, a letter would arrive here in a few days. This means there was no fire. In the evening of that day (09/02), a rather strange cloud appeared in the west-north and west-south-south. Although some said that it was the glow of the sunset in the sky that had persisted, it was quite strange that this glow was visible even at night. According to the authority at *Hirosaki*, the sky had gotten red before a severe earthquake and it was not a fire. Maybe a fire broke out in Russia. Some said it might have broken out at *Soya* or *Karafuto* (Sakhalin). I could not understand what it was. I kept my eyes on the glow in the sky southward during the sunset on that day. The sky was as red as in no other case in these years.

**Location:** *Hirosaki* (弘前), *Aomori*, Japan (N 40°36′, E 140°28′)

**J4:** KNT: p315

**Original Text:** 安政六年八月六日，暮六つ時より戌の方より寅の方までかかり，火事の如くの雲焼，火事と見る人も有

**Translation:** On 1859/09/02, From *Kure Mutsu* (17:00-19:00), a burning cloud that resembled a conflagration was suspended from *Inu* (300°, north-west) to *Tora* (60°, north-east) and some people thought that it was a conflagration.

**Location:** *Hiraka* (平鹿町), *Akita*, Japan (N 40°36′, E 140°31′)

### 3.2 Weather during 1859/08/28−09/04

As for the contemporary weather, we found several diaries with weather descriptions: one in Korea and 12 in Japan. The observation points are as follows; Seoul (in Korea), Hirosaki, Tama, Akita, Odate, Shinjo, Kudoyama, Yamanashi, Yokohama, Hachioji, Kagoshima, Kawasaki, and Kyoto (in Japan). Their results are given in Table 2.

## 4. Discussion

### 4.1 Reliability of the records

In order to evaluate the reliability of the historical records for the Carrington event, it is crucially important to examine the nature and the origin of the source documents. Note that by "reliability" we do not mean the likeliness of the records being actually those of aurora. Firstly, we evaluate the reliability by examining whether the records have a possibility to be documented by the observer him/herself, hearsay, or citation from the older documents. Secondly, we evaluate the reliability of each source document from the perspective of accessibility to the original version, namely whether the very first manuscript by the author is available or only transcriptions or citations are available. From this perspective, the closer to the original, the more reliable the document is.

Document LZ (that includes record C1) is a local treatise for *Luánchéng* in China compiled during 1872−74. This record is a compilation and includes a chapter for omens (祥異) since the very early dynasty in China. Considering the date on which the Carrington event occurred, which is less than 20 years prior to its compilation, there is a good possibility that the author himself or a close person witnessed the event itself.

KTN (J1) seems to be a chronicle compiled based on the *Kumano Nendaiki* (熊野年代記) in the Meiji era. The *Kumano Nendaiki* manuscript has two versions: the *Umemoto* Manuscript (梅本本), which was compiled by the hermit lord of *Shingu,* and the *Ono* Manuscript (小野本), which is a copy of the former in the Meiji era. Although the *Umemoto* Manuscript involves the history of *Shingu* up to 1868, it does not include the description of the Carrington event. The *Ono* Manuscript involves history only up to 1727 and naturally does not include the records in question. Therefore, its record J1 remains only in KTN cited in *Shingu* City (1935), which seems puzzling. Although we could not find the original manuscript, KTN itself seems to have been compiled in the Meiji era. As its name suggests, this document is a compilation of the records in *Kumano* district although there is a possibility that the last author personally witnessed the event itself, considering that the Carrington event happened less than 9 years before the end of this chronicle.

YUS (J2) is a diary by *Ubei*, a head of *Yamaguchi* (山口) Village in *In'nan* (印南). It seems to have been composed of seven volumes, although the first one is lost. It is the second volume (1849−1859/11) that contains records of the Carrington event. This diary involves various topics, such as matters of *Ubei*'s family, management of the village, or rumors. The fact that not every date had an entry suggests the possibility that this diary was compiled after the events themselves had occurred. *Ubei* himself engaged in various businesses such as rice trade and shipping. In'nan-cho (1987) respects the original form of historical documents and involves the full text of historical documents when showing the sources. Therefore, we can regard this book as a source document. The location of the original manuscript is unknown as its owner left this town, according to the education committee in In'nan town.

YKM (J3) is a diary of a merchant *Matasaburo Kanagiya* (金木屋又三郎) in 21 volumes. His family was originally a local merchant family from a rural village distant from the *Hirosaki* Castle. After becoming rich by engaging in brewing and pawnbroking, they came to *Hirosaki* to become official merchants in collaboration with *Daidoji* (大道寺) Family, an aristocratic family there. Although they perished in the turmoil during the Meiji Revolution and their documents were lost, their diary was later found with an antiquarian and bought by *Hirosaki* City. One of the most important aspects of this diary is that it contains detailed records of weather and disaster not only in *Hirosaki* but also in other areas such as *Edo* in detail. He kept his record for weather every 2−3 hours. This is good evidence that he wrote his diary every day. This respect makes his diary quite unique and ascertains quite high reliability of his diary.

KNT (J4) is cited in Hiraka-cho (1969). We presume that this is a diary or chronicle by a village head. The location of the original manuscript is unknown today and this derivative document was in a private collection (Watanabe 2007).

From our second perspective, C1 and J3 are the most reliable because their original manuscripts are available. J2 is reliable as its critical edition is cited precisely in In'nan-cho (1976), although we cannot determine the present location of the original manuscript. J1 is relatively suspicious because we have no information for the original source of the record, especially that for 1859. As for J4, we cannot assign much credibility to it because we have only a passage cited in Hiraka-cho (1969).

**4.2 Observational Sites**

All the East Asian aurora records presented in this paper was on 09/02, when the aurora was witnessed down to 20° N geomagnetic latitude (Loomis 1859-62; Kimball 1960; Green & Boardsen 2006). The geomagnetic latitude for all the records presented in this paper is calculated and shown in Table 1. All of them are as low as 20−30° N.

The geomagnetic latitude is defined as the angle between a point of observation and the magnetic equator. The location of the magnetic pole was calculated using the spherical harmonic coefficients provided by the gufm1 model (Jackson et al. 2000). As for the points listed in Table 1, the difference between the magnetic latitude derived from the gufm1 model, and that derived from the IGRF 1900 model (Thébault et al. 2015), is less than 0.25 degrees.

In China, we have no reports from official documents such as QQ but one from local history (LZ) at *Luánchéng*. We are not very sure if the aurora was seen on this date as Chinese astronomers only recorded events of significance in the historical documents, as is seen in *Míngshǐ* (Astronomy III, p 417).

As we have already mentioned above, Korean official records such as ISN (pp834−847) or SCWIG (pp397−408) do not include any aurora-like events around this time. Reviewing reports of the weather, we found that it was sunny on 09/02 but rainy on 09/01 and 09/03. Therefore, we can assume that the weather was not stable, and that the clouds bringing rain on 09/03 could have obscured the sky at night on 09/02.

Japan has two groups of observation points, Wakayama prefecture (J1, J2) and the northern most of Honshu Island (J3, J4), without reports from the central latitude of Japan. The weather on 1859/09/02 itself was not bad as is seen in Table 2, contrary to speculation by Green et al. (2005). It was generally sunny except for *Kyoto*, *Kanto* region, *Kagoshima*, and *Kanzawa* (JW04, JW10, JW13, JW14, JW15, JW16); although in another record the weather in *Kanto* was also reported to be sunny (JW11). What attract our interest are reports (JW15, JW16) that the sky got cloudy during the evening at *Hachioji* and at *Kanazawa*. The weather records on the following date 09/03 are generally the same with the ones on 09/02 except for several sites: rainy in *Tama* (JW04), cloudy in *Shinjo* (JW07), cloudy in Yokohama (JW10) and rainy in Kagoshima (JW13). These reports show a possibility that a band of clouds appeared in the central latitude of Japan at night on 1859/09/02 that prevented observation of the aurora.

All the observational sites are shown in the map of East Asia in Table 1 and Figure 7.

**Aurora Oval**

These historical records have shown that the aurora was seen at the geomagnetic latitude greater than ~23 degrees in the East Asia's meridian. In the America's meridian, the lowest latitude at which the aurora was seen was ~22 degrees at 05-10 UT on September 2 (Kimball 1960; Tsurutani et al. 2003; Green & Boardsen 2006). Thus, the lowest latitude at which the aurora was seen in Japan (~23 degrees) was comparable to that in the America's meridian (~22 degrees). During magnetic storms the plasmasheet comes inward from the tail. The innermost location will be the plasmapause (the boundary of the plasmasphere) and it may be there where the red auroras would be seen (Tsurutani et al.2003). Yokoyama et al. (1998) investigated an empirical relation between the equatorward edge of the electron plasma sheet and the maximum accumulated particle energy in the inner magnetosphere, and thus, the Dst index: $Dst \approx -3400\cos^6 \lambda + 60 \text{ nT}$, where $\lambda$ is the magnetic latitude at the surface of the Earth. All the relevant Japanese records compare this aurora with conflagration at some distance. These descriptions remind us red auroras observed near the horizon. Assuming that the aurora was seen near the horizon and that the equatorward edge of the aurora was located at 27 degrees, the estimated Dst is at least -1640 nT. This is comparable that estimated by Tsurutani et al. (2003), but we avoid discussing the estimated Dst because of the difficulty in evaluating single stations.

**Time of aurora observation in East Asia**

We compare the timing of the East Asian observations with the magnetic field measurement at Bombay presented in Tsurutani et al. (2003). It should be noted here that the contemporary East Asian time (不定時法; or varying length system), which divides seasonally changing day/night into 6 equal parts with average length of approximately 2 hours. The time is expressed by the name for each part. Thus, unless explained explicitly, it is not possible to tell whether the observed phenomena were seen at an instance or kept visible during the ~2 hours. In this paper we write the possible range of time when we translate the Japanese time to English.

In the record J3-1, the aurora was observed in the morning (approximately 05:00–06:00 Japan Local Time (JLT; UT+9h) —21:00–22:00 UT—on 09/01) on 09/02 in J3-1. In J2, it starts by describing the red color of clouds at about 15:00–17:00 JLT (06:00–08:00 UT) but the main aurora-like descriptions were in 21:00–23:00 JLT (12:00-14:00 UT). The other records ranges from 18:00 JLT (09:00UT) to 06:00 JLT (21:00UT).

On the other hand, the Bombay magnetogram examined by Tsurutani et al. (2003) shows two peaks caused by the Carrington event and second storm after the Carrington event: the first at 05:30–6:00 (UT) and the second at 14:30 (UT). The first one corresponds with 12:00–12:30 in Bombay local time (BLT; UT+5.5) and 15:30–16:00 JST (UT+9). The latter corresponds with 20:00BLT (UT+5.5) and 23:30 JST (UT+9).

Figure 8 shows the data compared between both the Bombay magnetogram (Tsurutani et al. 2003) and the time of aurora observation in East Asia on 1859/09/02. Yokoyama et al. (1998) indicates that the equatorward boundary of the aurora belt appeared to reach the lowest latitude 0±2 hours before Dst index reached its peak, although the deviation is higher for more intense storms. This may reflect the fact that the plasma sheet electrons are transported earthward by the enhanced convection electric field. As the plasma sheet electrons move earthward, the equatorward edge of the electron plasma sheet, that is, the auroral oval moves equatorward. This period corresponds to the storm main phase. When the magnetogram showed the largest excursion in Bombay, the aurora was not observed in Japan because of sunlight. A few hours later, the aurora started to be recorded in Japan. This period includes the recovery phase of the storm. This is consistent with the modern observational result that low-latitude aurora appears not only in the storm main phase, but also in the storm recovery phase (Shiokawa et al., 1997; Shiokawa et al., 2005).

The time of J3-1 roughly corresponds with the latitude envelop of the aurora at midnight between 09/01 and 09/02, shown in Figure 3 by Green & Boardsen (2006). The times of the rest of the records roughly correspond to the second peak at 23:30 JST, shown in Bombay magnetogram according to Tsurutani et al. (2003)[3].

---

[3] In this time, it was not aurora by human eyes but only magnetic storm by magnetogram that was observed at Bombay. However, in 1872, another magnetic storm caused aurora observation even at Bombay (Chapman and Bartels 1940; Tsurutani et al. 2005). This is another example of great

What is mysterious is J2. This record presents us "a red cloud" in the north at 15:00–17:00 JST, about two hours after the magnetogram peak. Considering the date of observation, it is well before the sunset, and hence not likely to be the aurora itself. However, it is interesting to note the report from Halifax (44°39′ N) on 08/28. It was included in Loomis (1860b: 251) that cites a letter by Lieut. N. Home which described the auroral observation there as follows:

> 08/28 17:00, I noticed a long narrow belt of cloud from E. to W. having a peculiar orange-white appearance.
>
> 08/28 20:00, I observed this cloud (which in the interim appearance to be stationary) suddenly became luminous at its eastern extremity. The cloud was 10° wide, and appeared to extend from horizon to horizon; no other clouds were visible.

It seems likely that this "long narrow belt of cloud" was not a normal "cloud" as it got suddenly brighter at 20:00. The description for the observation at 20:00 in Loomis et al. (1860b) reminds us of the aurora is associated with substorm-associated activity (Shiokawa et al. 1994). However, considering the date of the observation; 17:00, when the "belt" itself was noticed, seems to be before sunset. Although there are daytime auroral observations using spectrometers (e.g., Rees et al. 2000; Pallamraju & Chakrabarti 2005), it seems very unlikely that the aurora was visible to the naked eyes. However, considering the extraordinary nature of the Carrington event, we leave open the possibility of twilight or daytime auroral observations (see, Loomis 1960b). How intense the auroral emission could be is indeed an interesting theoretical problem, particularly in the context of exoplanets, where the detection of the auroral emission from atomic oxygen could be considered evidence of the existence of oxygen and hence, an indicator of the possibility of biological processes on the planet (Akasofu 1999).

In the mean time, our results also present us insights into how auroras during strong magnetic storms are described by the ancient observers before the period of modern observations. This is of great importance for the investigation of pre-modern aurora-like records as pre-modern observers did not understand the nature of aurora and sometimes included other phenomena such as atmospheric optics or comets as is shown in Keimatsu (1970), Hayakawa et al. (2015), or Chapman et al. (2015).

---

magnetic storms to be discussed.

**Conclusions**

We surveyed auroral records in East Asia around the time of the Carrington event and examined one record from China and four records from Japan, including some newly found ones. These records remedy the previous lack of observation in East Asia. Our results provide clear, concrete examples of pre-modern oriental records for extreme space weather attested by modern observation for the same time. They present us a clear insight into investigation of pre-modern observations of extreme space weather events (before the age of modern observation) by astronomers or historians who did not have a concrete understanding of the nature of auroras.

**Acknowledgements**

We acknowledge support by the Center for the Promotion of Integrated Sciences (CPIS) of SOKENDAI as well as the Kyoto University's Supporting Program for Interaction-based Initiative Team Studies "Integrated study on human in space" (PI: H. Isobe), the Interdisciplinary Research Idea contest 2014 by the Center of Promotion Interdisciplinary Education and Research, the "UCHUGAKU" project of the Unit of Synergetic Studies for Space, and the Exploratory and Mission Research Projects of the Research Institute of Sustainable Humanosphere, Kyoto University. This work was also encouraged by Grant-in-Aid from the Ministry of Education, Culture, Sports, Science and Technology of Japan, Grant Number 15H05816 (PI: S. Yoden), 15H03732 (PI: Y. Ebihara), and 15H05815 (PI: Y. Miyoshi). We thank Dr. T. Takeda for his fruitful advices and assistances. We also thank Hirosaki City Library for their permission to use the pictures of manuscript of YKM.

| ID | Year | Month | Date | Color | Description | Direction | Start | End | Place | G. Latitude | G. Longitude | GM Latitude |
|---|---|---|---|---|---|---|---|---|---|---|---|---|
| C1 | 1859 | 9 | 2 | R | V | wn-en | | Dawn | *Luánchéng*, China | N37°54′ | E114°39′ | 26.518 |
| J1 | 1859 | 9 | 2 | R | | n | 17:00-19:00 | Midnight | *Shingu*, Japan | N33°44′ | E135°59′ | 23.077 |
| J2 | 1859 | 9 | 2 | R | C | n | 15:00-17:00 | 21:00-23:00 | *In'nan*, Japan | N33°49′ | E135°39′ | 23.139 |
| J3-1 | 1859 | 9 | 2 | R | | wn-en | 05:00-06:00 | 05:00-06:00 | *Hirosaki*, Japan | N40°36′ | E140°28′ | 30.237 |
| J3-2 | 1859 | 9 | 2 | R | | n | | | *Hirosaki*, Japan | N40°36′ | E140°28′ | 30.237 |
| J3-3 | 1859 | 9 | 2 | R | | n | | | *Hirosaki*, Japan | N40°36′ | E140°28′ | 30.237 |
| J4 | 1859 | 9 | 2 | fire | C | wn-en | 17:00-19:00 | | *Hiraka*, Japan | N39°12′ | E140°34′ | 28.852 |

Table 1: Records of aurora observation caused by Carrington flare.

| Date | 08/28 | 08/29 | 08/30 | 08/31 | 09/01 | 09/02 | 09/03 | 09/04 | Reference |
|---|---|---|---|---|---|---|---|---|---|
| Hirosaki | clear | clear | clear | cloudy | cloudy rainy | sunny | sunny | sunny | JW03 |
| Tama | clear | clear | clear | good | good | cloudy | rainy cloudy | good | JW04 |
| Akita | cloudy sunny | cloudy | xx | sunny/rainy | sunny/rainy | rainy sunny | sunny rainy | sunny | JW05 |
| Odate | cold (morning) | cloudy | good | good | rainy | good | good | hot | JW06 |
| Shinjo | good | good | good | good | good | good | cloudy | good | JW07 |
| Kudoyama | sunny | sunny | clear | sunny | sunny | clear | sunny | sunny | JW08 |
| Yamanashi | good | good | good | good cloudy | good | good | good | good | JW09 |
| Yokohama | sunny | sunny | sunny | sunny | sunny | cloudy | cloudy | sunny | JW10 |
| Hachioji | sunny | sunny | sunny | sunny | sunny | sunny | sunny | sunny | JW11 |
| Kawasaki | hot | hot | cloudy hot | hot thunder | hot | hot | xx | hot | JW12 |
| Kagoshima | sunny | sunny | sunny | cloudy sunny | cloudy | cloudy sunny | rainy sunny | rainy windy | JW13 |
| Kyoto | sunny | sunny | sunny | sunny | sunny | cloudy | sunny | sunny | JW14 |
| Hachioji | sunny | sunny | good | good | sunny | good rainy | rainy | good | JW15 |
| Kanazawa | sunny | sunny | sunny | sunny | clear | rainy cloudy | sunny | sunny | JW16 |
| Seoul | xx | rainy | xx | rainy | rainy | xx | rainy | xx | ISN, pp834-847 |

| Seoul | sunny | rainy | sunny | rainy | rainy | sunny | rainy | sunny | SCWIG, pp397-408 |

Table 2: Weather in East Asia from 1859/08/28 to 1859/09/02.

Abbreviations: clear (快晴), sunny (晴), good (天気, 天気よし), cloudy (曇, 陰), rainy (雨), xx (lack of record)

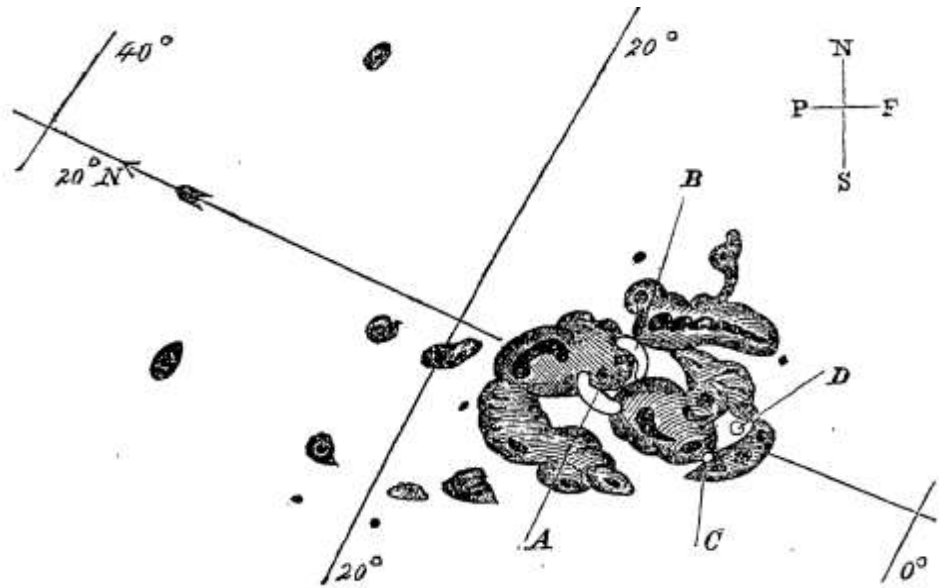

Figure 1: A sketch of sunspots by Carrington (1859, p13)

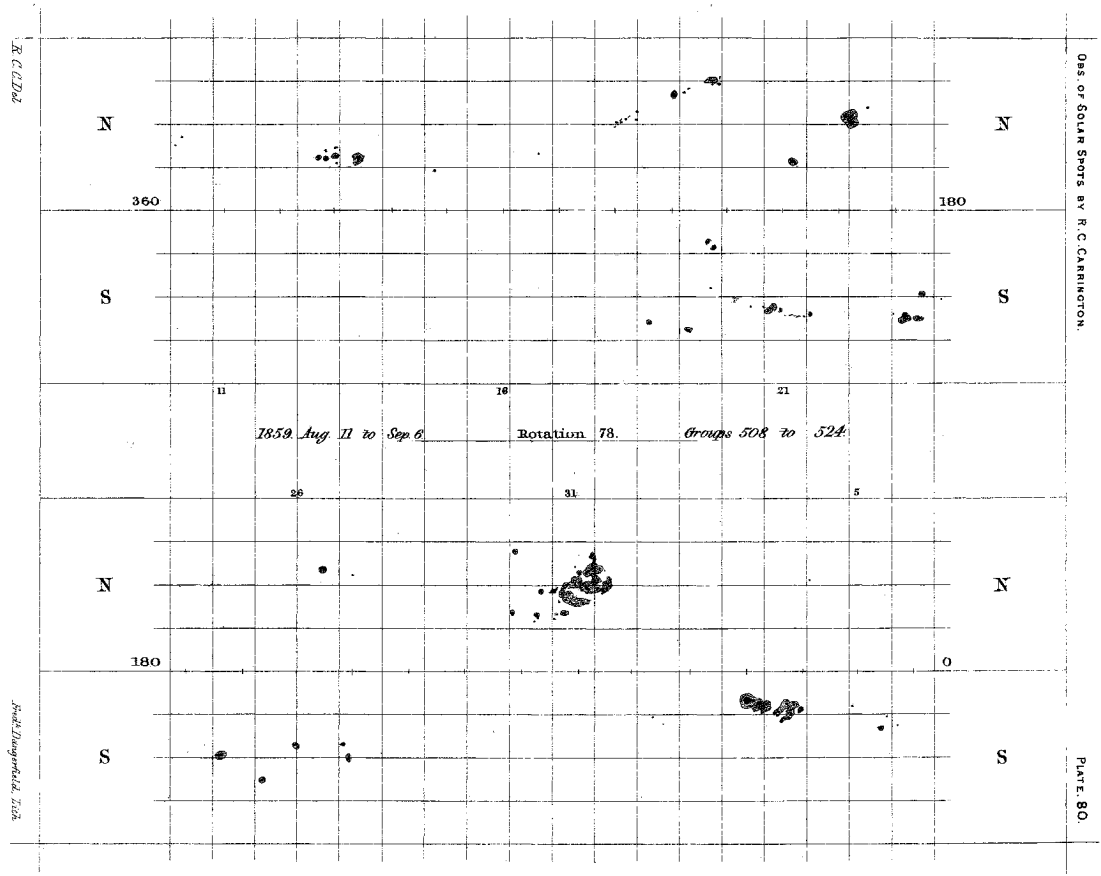

Figure 2: A sketch of sunspots with whole solar disc by Carrington (1863, plate 80)

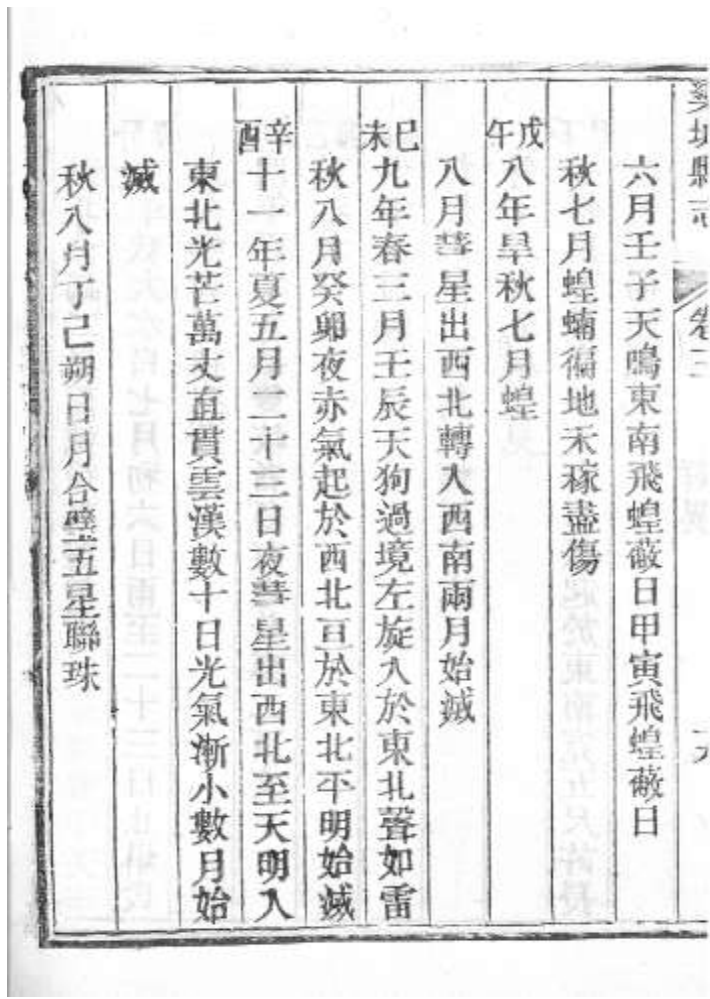

Figure 3: Picture of Manuscript of C1 (with permission from Institute for Research in Humanities)

Figure 4: Picture of manuscript of J3-1 (left end)

Figure 5: Picture of manuscript of J3-2 (right end)

Figure 6-1: Picture of manuscript of J3-3

Figure 6-2: Picture of manuscript of J3-3 (continued)

Due to the matter of license, we cannot show these figures on the preprint version.

Please see the published version in PASJ for these figures.

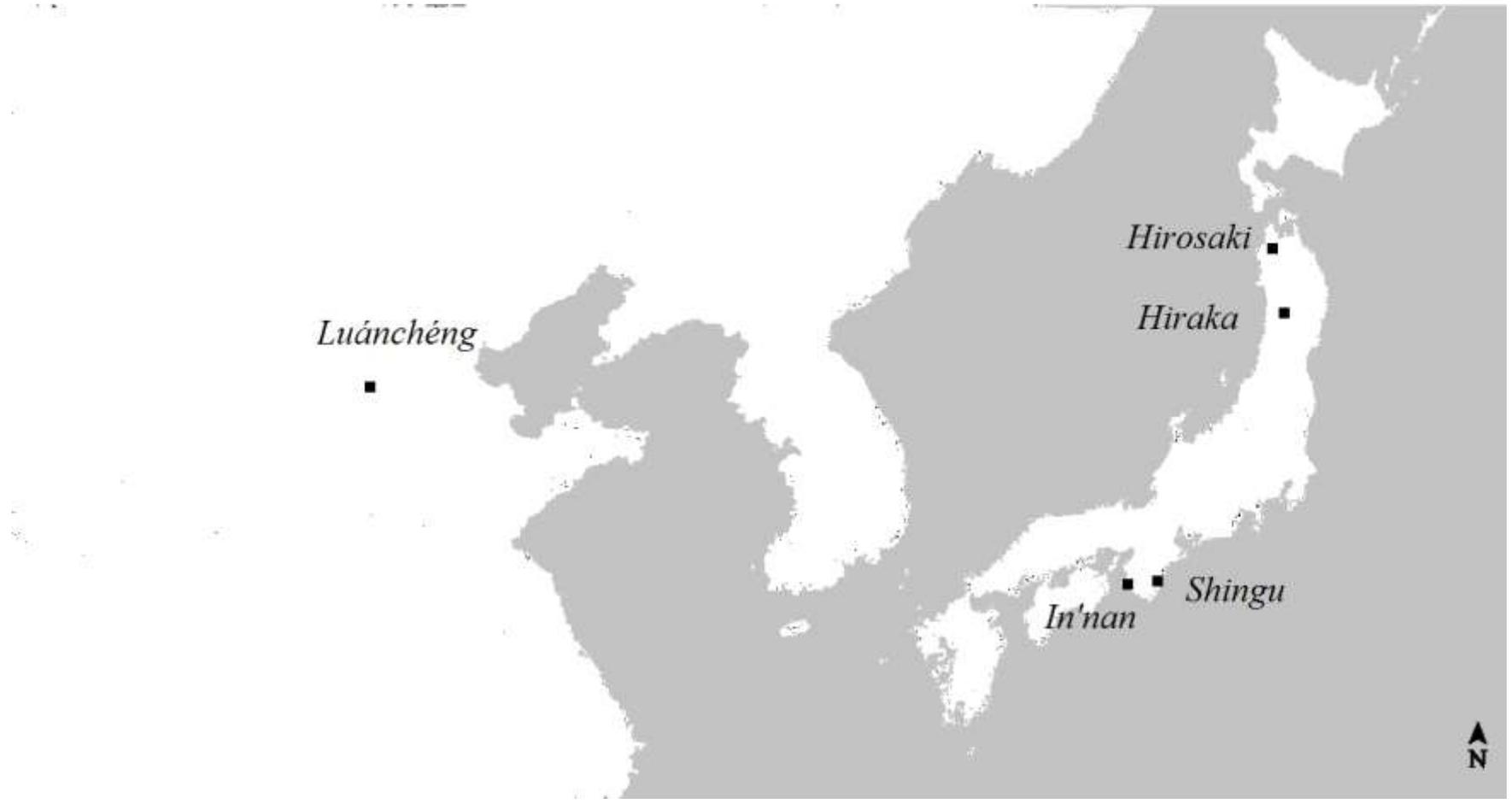

Figure 7: Map of observational sites of auroras caused by Carrington Event.

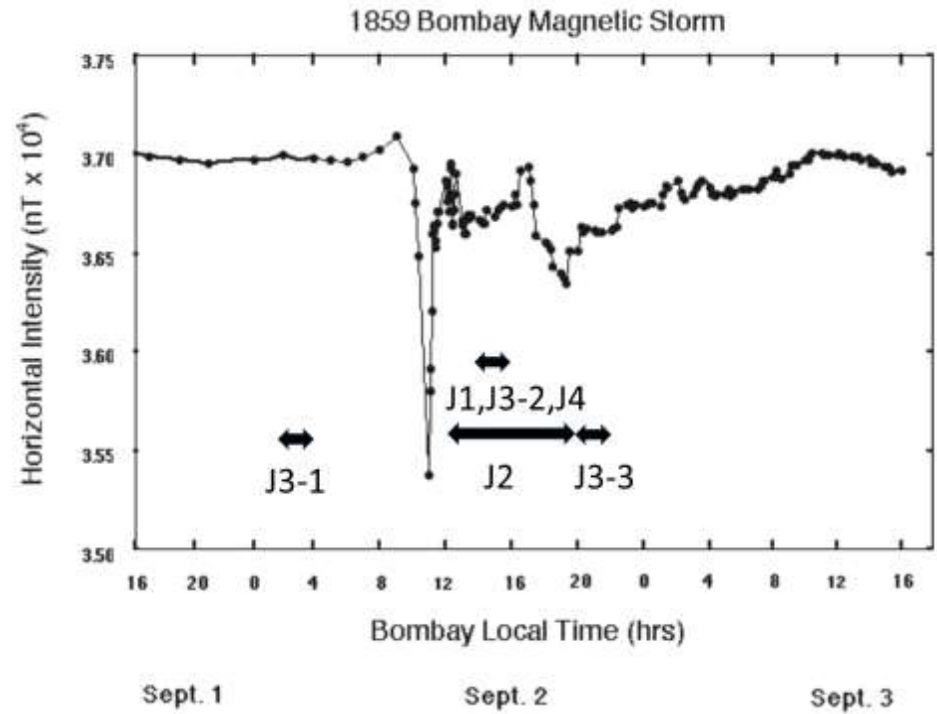

Figure 8: Magnetogram of the horizontal geomagnetic field intensity from Bombay, India, covering the magnetic storm of September 2, 1859. Adopted and modified from Tsurutani et al. (2003). Timings of aurora observations in Japan are also indicated.